\documentclass[aps,prd,onecolumn,12pt,preprintnumbers]{revtex4-1}

\usepackage{hyperref}

\usepackage{float}

\usepackage{tikz}

\usepackage{graphicx}
\usepackage{epstopdf}

\usepackage{caption}

\begin{document}

\preprint{HDP: 20 -- 01}

\title{Banjo Break Angle Tension Modulation as Parametric Oscillation}

\author{David Politzer}

%\email[]{politzer@theory.caltech.edu}
\email[]{politzer@theory.caltech.edu}

\homepage[]{http://www.its.caltech.edu/~politzer}

%\thanks{452-48 Caltech, Pasadena CA 91125}
\altaffiliation{\footnotesize Pasadena CA 91125}
%\altaffiliation{\newline \em \em \em 452-48 Caltech, Pasadena CA 91125}
\affiliation{}

%\date{\today}
\date{January 9, 2020}

\begin{figure}[h!]
\includegraphics[width=3.0in]{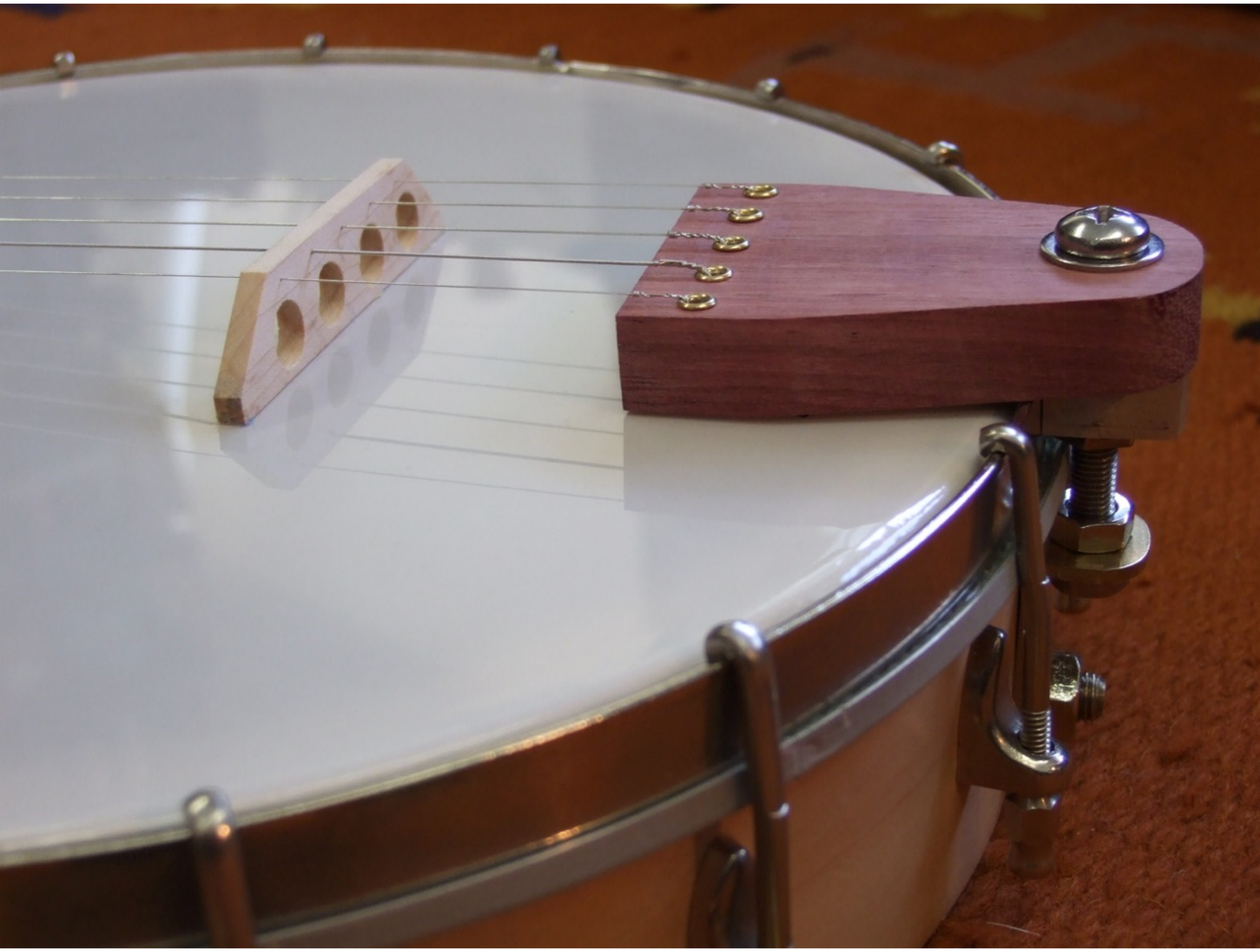}
\includegraphics[width=3.0in]{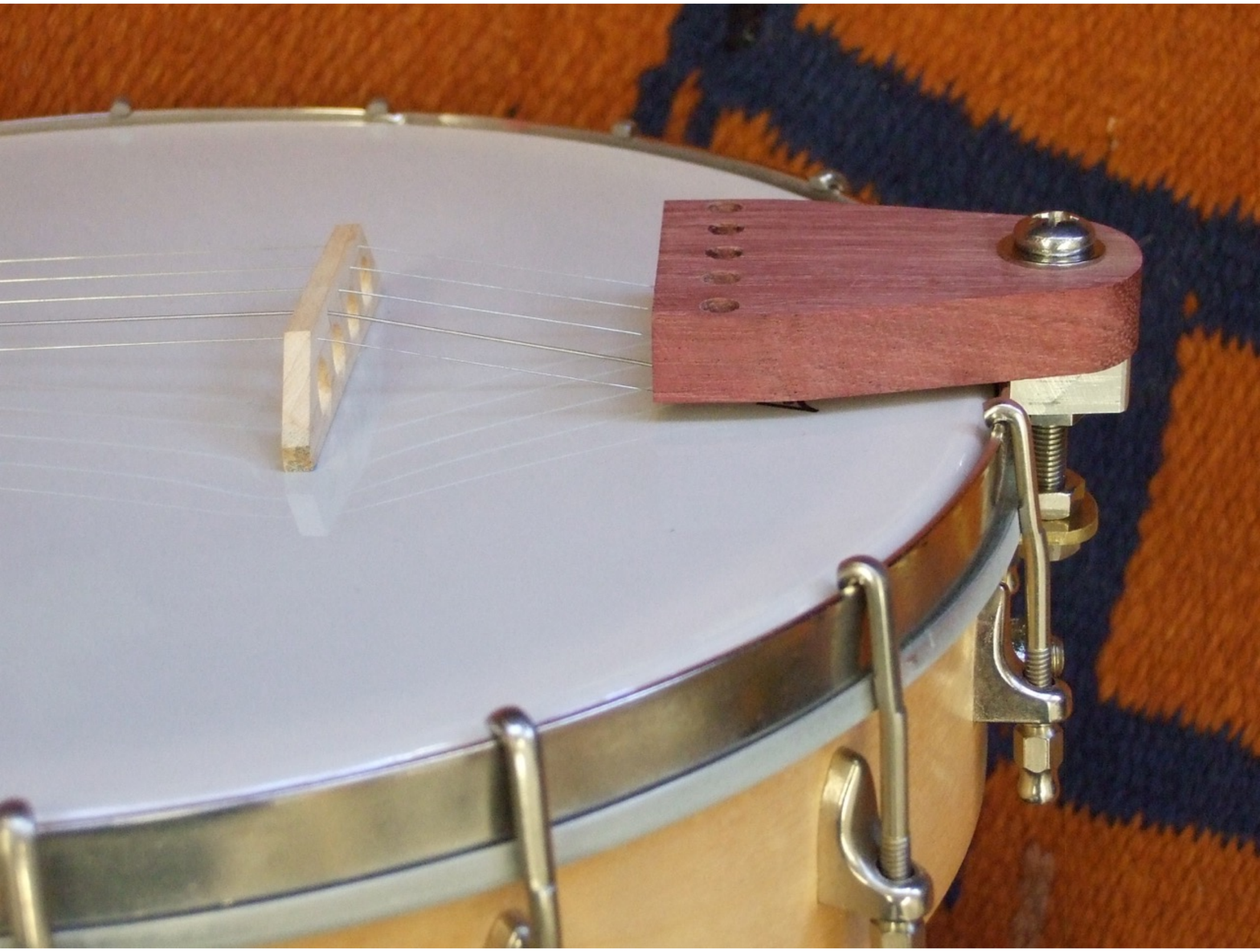}
\end{figure}

\bigskip

\begin{abstract}
The motion of the floating bridge of the banjo, in conjunction with the break angle of the strings over that bridge, produces string tension modulation that is first order in the amplitude of the string motion.  This note refines a previous suggestion regarding the impact on the frequencies of the strings' and bridge's motion.  For a given mode frequency pair of string and bridge, the resulting tension modulation produces a new, additional motion characterized by the sum and difference of the original ones.  Strictly speaking, this corresponds to canonical ``frequency modulation" only in the limit of modulation slow compared to the string frequency.  The more general result is precisely an example of what is known as ``parametric oscillation," first analyzed by Rayleigh.  The qualitative impact of tension modulation on banjo timbre remains as suggested previously.  It is only the precise math and physics that warrants this correction.

\end{abstract} 

\maketitle{\centerline {\large \bf Banjo Break Angle Tension Modulation as Parametric Oscillation}}

\bigskip

\bigskip

\centerline{\bf BACKGROUND}

\medskip

As they pass over the bridge, the strings of the banjo form a ``break" angle --- typically somewhere between $3^{\text{o}}$ and $16^{\text{o}}$ --- which is determined by the height of the bridge and the geometry of the tailpiece, as illustrated in FIG.~1.

\begin{figure}[h!]
\includegraphics[width=3.8in]{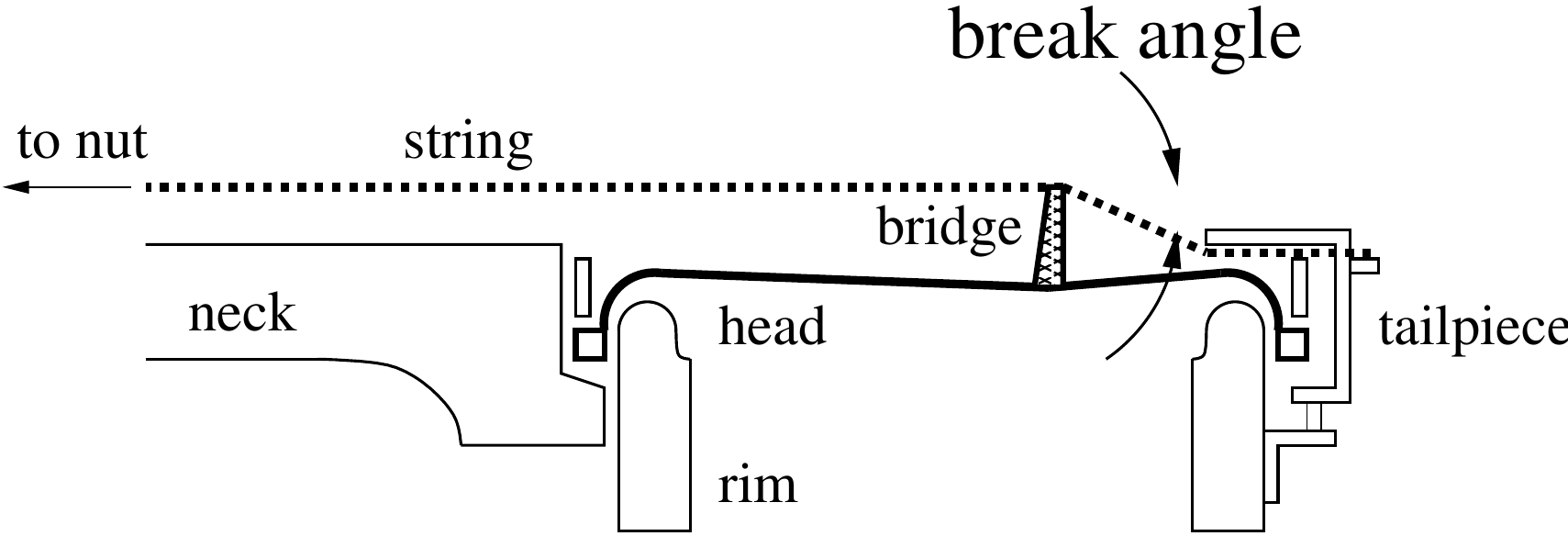}
\caption{ }
\end{figure}  It is long-known to players and builders that increasing that angle enhances the ``ring" and ``clang" of the instrument.  A direct demonstration of that phenomenon was presented in ref.~\cite{zero}, which contains sound samples of a particular instrument where the break angle can be either zero or $13^{\text{o}}$.  Note that zero degrees requires a specially designed bridge to allow the strings to function normally without buzzing, as shown on page 1.  Links to the recordings themselves are also given in this note's appendix.

For a given length and linear density, the fundamental pitch frequency of a string is proportional to the square-root of the tension.  It was an obvious leap to imagine that tension modulation produced frequency modulation.  The sound of audio range frequency modulation has a rich and wonderful history since its 1973 discovery by John Chowning, the licensing to Yamaha to become Stanford University's second highest earning patent, the production of the Yamaha DX7, and all that followed.\cite{chowning}  Chowning described his first examples as sounding metallic.

While the string stretch associated with small vibrations and fixed ends is second order in the amplitude, a non-zero break angle produces stretching that is first order in the amplitude when the bridge moves.  A natural assumption was that this phenomenon produced the sound of Chowning's frequency modulation.\cite{FM}

But frequency modulation is usually understood to have a precise mathematical meaning.  An initially sinusoidal oscillation in time is modulated in frequency at some other sinusoidal frequency and amplitude.  When the modulation is weak in amplitude and much lower in frequency, its principal effect is to add side bands at the sum and difference frequencies.  (This is the  basis of FM radio transmission.)

An explicit analysis of the basic equations of motion given below shows that the physical mechanism of tension modulation literally produces frequency modulation only in the limit that the modulation frequency is much less than the initial string frequency.  In the case of the banjo, there are certainly frequency components of the tension modulations that are comparable and higher than some of the initial string frequencies.  And then a purely sinusoidal tension modulation does {\it not} translate into a purely sinusoidal frequency modulation.  What does in fact happen is an example of what has come to be called parametric oscillation, which was first analyzed by Rayleigh.\cite{rayleigh}  For a given modal frequency pair of string and small amplitude bridge motion, the result is new, additional string and bridge motions characterized by the sum and difference frequencies of the original ones.

\bigskip

\centerline{\bf TENSION MODULATION IN THE STRING EQUATION OF MOTION}

\medskip

The zeroth order string motion can be expressed as a sum over the ideal normal modes.  It could be just one of them or plucked triangular traveling waves or any other possible free motion of the string.  The tension modulation (again, assumed to be simultaneous along the length of the string) is a function of time.  It need not have the periodicity of the string or even be periodic.  It is the net result of the forces of all modes of all the strings on the bridge and head.  In any case, it has a Fourier transform.  Even if there is only a single string's motion determining the bridge motion, in Fourier space we generally have to consider terms with string and bridge having different frequencies.

In the following, a dimensionless parameter $b$ is the relative amplitude of the tension modulation and is assumed to be small.  As shown in ref.~\cite{FM}, $b$ depends linearly on the string amplitude, $A$.  I work to lowest non-trivial order in $b$ because keeping anything higher than $b^1$ would be inconsistent --- such things having been ignored in the ideal string equation in the first place.  

So (as they say) without  loss of generality, we can consider a single ideal string mode, with frequency $\omega_o$ and wave number $k$, as the zeroth order string motion, and a single frequency of tension modulation $\omega$.  In particular, let the tension $T(t)$ be $T_o (1+b\text{ cos }\omega t)$.  
Let $\psi(z,t)$ be the string transverse displacement.  For small amplitudes, it satisfies 

\centerline{$\rho \ddot \psi = T(t) \psi''$ .}

\noindent And let $\psi = \psi_o + b \phi$ , with both $\psi_o$ and $\phi$ vanishing at both ends.  Take

\centerline{$\psi_o = A \text{ cos }\omega_o t \text{ sin }kz$ .}

Even before trying to find $\phi$, the zeroth order $\psi_o$ exerts an ${\cal O}(b)$ vertical force on the bridge of $b T_o \text{cos }\omega t {\partial \over {\partial z}}\psi_o$ = $b T_o A k \text{ cos }\omega t  \text{ cos }\omega_o t =  {1\over2}bT_oAk(\text{cos }\omega_+t + \omega_-t)$, where $\omega_\pm \equiv \omega_o \pm \omega$ . 

To find the equation for $\phi(z,t)$ to lowest order in $b$:  Let $v_o \equiv  (T_o / \rho)^{1/2} = \omega_o / k$ .  Keeping only the terms linear in $b$, you get

\centerline{$\ddot \phi = v_o^2 \phi'' + v_o^2 \text{ cos }\omega t$ $\psi_o''$ .}

Let $\phi(z,t) = f(t) \text{ sin } kz$.  Then

\centerline{$\ddot f(t) = -v_o^2 k^2 f(t)  - v_o^2 k^2 A \text{ cos } \omega t \text{ cos } \omega_o t = -\omega_o^2 f(t) - {1 \over  2} \omega_o^2 A (\text{cos }\omega_+ t + \text{cos }\omega_- t)$ .}

\noindent  So $f(t)$ is simply a forced SHO.  You add the particular solutions for the two forcing frequencies.  %Note that the only way $\omega_o = \omega_\pm$ is if $\omega = 0$, which is just a permanent retuning of the string.  
In particular, with 

\centerline{$f(t) = f_+(t) + f_-(t)$ ,}

\centerline{$f_\pm(t) = {{\omega_o^2 A} \over {2(\omega_o^2 - \omega_\pm^2)}} \text{ cos } \omega_\pm t =  {{\omega_o^2 A} \over {2 \omega(\mp 2\omega_o - \omega)}} \text{ cos } \omega_\pm t$ .}

\medskip

\noindent The general solution for $f(t)$ also includes oscillation at $\omega_o$, independent of $\omega$.  That part should be absorbed in $\psi_o(t)$.  The pole at $\omega = 0$ is just a permanent retuning of the string.  The pole in $f_-(t)$ at $\omega = 2 \omega_o$ is the standard parametric oscillator resonance.  In particular, tension modulation at $\omega \approx 2\omega_o$ drives the string mode with $\text{cos }\omega_- t \approx \text{cos }\omega_o t$, i.e., approximately at the original mode frequency.

The resulting vertical force on the bridge due to $\phi(z,t)$ is $b T_o k f(t)$.  Adding this to the ${\cal O}(b)$ force on the bridge due to $\psi_o$ and $T(t)$, the total ${\cal O}(b)$ force on the bridge is
\centerline{${1 \over 2} T_o b A  \{(1 - {1 \over {{\omega \over \omega_o}(2+{\omega \over \omega_o})}}) \text{cos}(\omega_o + \omega)t + (1 + {1 \over {{\omega \over \omega_o}(2-{\omega \over \omega_o})}}) \text{cos}(\omega_o - \omega)t \}$ .}

\bigskip

\centerline{\bf DECAY TIMES \& RESONANCE WIDTHS}

\medskip

$A$ is the amplitude of a plucked string mode.  So it decays in time more or less (or exactly) like an exponential.  The modulation amplitude $b$ also decays similarly, although potentially with a somewhat different decay time.  (Strictly speaking, the decays can be the sum of more than one exponential and also may exhibit beats.)  The tension-modulation-induced amplitude decays like $bA$, i.e., faster than either separately.  Hence, the tension modulation effect appears in the attack or early part of the pluck sound.

These decay times are also reflected in the behavior near the resonance in $f_\pm(t)$.  The actual behavior will have a finite maximum and a related resonance width.  This could be exhibited explicitly by including damping in the equations of motion.  But it is also a necessary consequence of the fact that the resonance amplitude can only build up to a finite value in a finite amount of time.

\bigskip

\centerline{\bf \large THE PROBLEM WITH FREQUENCY MODULATION}

\medskip

The starting equation, $\rho \ddot \psi = T(t) \psi'' = T_o(1+b\text{ cos }\omega t)\psi''$, suggests that the whole effect might simply be viewed as a modulation of $v_o$, i.e., $v_o \to v(t) = v_o(1 + {1 \over 2}b\text{ cos }\omega t)$ or a modulation of the flow of time, e.g., $t \to \tau(t) =  t - {b \over {2\omega}}\text{ sin } \omega t$.  However, any such ansatz for the wave $\psi$ does not satisfy the basic equation of motion.  The discrepancy is ${\cal O}({\omega \over \omega_o} b)$, i.e., it goes away if the modulation is slow compared to the original frequency.

In terms of traveling waves, tension modulation changes the speed of travel but not the wave shapes in space only if $\omega \ll \omega_o$.  More generally, tension modulation changes the wave shapes as well.

\bigskip

\centerline{\bf CONCLUSION}

\medskip

The implication of all this is that the general motion of the banjo floating bridge with a non-zero break angle can introduce components into the early part of the sound that are not the harmonics of the plucked string.  Note that even for frequencies already present in the zeroth order string motion, additional sounds at those same frequencies will alter the timbre.  I would welcome any suggestion of an alternative explanation for the differences in sound that are presented in the Appendix, below.  A proper analysis would require a model of the total bridge motion and a double sum over string and bridge motion modes.  The low frequency string modes generally produce the largest bridge motion ($b$), so it's expected that significant contributions come from $\omega \ll \omega_o$ and this gives the sound of canonical frequency modulation.  

\bigskip

\bigskip

\centerline{\bf APPENDIX}

\medskip

Here are links to the sounds of the $0^{\text{o}}$ and $13^{\text{o}}$ break angle configurations.  The banjo is a year 2000 Deering Goodtime, played with a solid disk back spaced $1/4''$ off the rim.  The head was tensioned to 91 on a DrumDial.  For two tunes, each played in the two configurations:

\href{http://www.its.caltech.edu/~politzer/zero-break/A.mp3}{click here for sample A}, 
 
\href{http://www.its.caltech.edu/~politzer/zero-break/B.mp3}{click here for sample B},
 
\href{http://www.its.caltech.edu/~politzer/zero-break/C.mp3}{click here for sample C}, and
 
\href{http://www.its.caltech.edu/~politzer/zero-break/D.mp3}{click here for sample D,}
 
\noindent or go to http://www.its.caltech.edu/\url{~}politzer/zero-break/A.mp3 and then switch A to B, C, and D.  If you can't hear which is which, this exercise is pretty bootless.

\end{document}